\documentclass[aps,superscriptaddress,twocolumn,amssymb]{revtex4-1}
\usepackage{revsymb4-1}
\usepackage{amsmath}
\usepackage{graphicx}
\usepackage{amssymb}
\usepackage{multirow}
\usepackage{float}
\usepackage{xcolor}
\usepackage{natbib}
\usepackage{booktabs}
\usepackage{filecontents}
\usepackage{chngcntr}
\usepackage[colorlinks=true, allcolors=blue]{hyperref}

\newcommand{\upperRomannumeral}[1]{\uppercase\expandafter{\romannumeral#1}}

\usepackage[normalem]{ulem}

\usepackage{siunitx}
\usepackage[version=3]{mhchem}

\begin{document}

\title{DMRG Analysis of Magnetic Order in the Zigzag Edges of Hexagonal \ce{CrN} Nanoribbons}

\author{Micha\l{} Kupczy\'{n}ski}
\affiliation{Institute of Theoretical Physics, Wroc\l{}aw University of Science and Technology, Wybrze\.{z}e Wyspia\'{n}skiego 27, 50-370 Wroc\l{}aw, Poland}
\author{Jaros\l{}aw Paw\l{}owski}
\email{jaroslaw.pawlowski@pwr.edu.pl}
\affiliation{Institute of Theoretical Physics, Wroc\l{}aw University of Science and Technology, Wybrze\.{z}e Wyspia\'{n}skiego 27, 50-370 Wroc\l{}aw, Poland}
\author{Aybey Mogulkoc}
\affiliation{Department of Physics, Faculty of Sciences, Ankara University, 06100 Tandogan, Ankara, Turkey}
\author{Mohsen Modarresi}
\email{m.modarresi@um.ac.ir}
\affiliation{Department of Physics, Faculty of Science, Ferdowsi University of Mashhad, Mashhad, Iran}

\date{\today}

\begin{abstract}
We investigate the finite temperature magnetic order at the edges of hexagonal \ce{CrN} nanoribbons by using the density-functional theory combined with the density-matrix renormalization group method. Moreover, the spin-dependent transport in nanoribbons is calculated within the semi-classical Boltzmann transport theory. We find out that the zigzag edges have lower energy with respect to armchair edges. The zigzag edge of \ce{CrN} nanoribbon shows half metallic electronic character which is the same as for the 2D monolayer. The localized electronic states on the zigzag edges reduce the electronic band gap energy for spin down electrons. The ab-initio electronic results are mapped into an effective 1D Heisenberg spin model up to the next nearest neighbor exchange interaction term. For zigzag ribbons, the nearest neighbor and next nearest neighbor magnetic exchange are around $10$ to $12$, and $-2$ to $0$ ~meV/Cr atom, respectively. The finite spin correlation length in 1D nanoribbons drops sharply to zero with temperature. The absence of long range spin correlations at the edges is a practical drawback for future room temperature 2D spintronic devices. The maximally localized Wannier functions are used for band interpolation and spin-dependent transport calculations by using the semi-classical Boltzmann equation. We show that zigzag edges of \ce{CrN} are perfect spin filter under both electron and hole doping.
%A 1D spin valve is proposed based on blue-phosphorus nanoribbon sandwiched between two zigzag \ce{CrN} nanoribbons. The electronic conductivity depends on relative direction of magnetic moments in these zigzag \ce{CrN} nanoribbons. For parallel alignment of magnetic moments in \ce{CrN} nanoribbons only spin up electrons contribute to conductivity. The magnetoresistance of \ce{CrN} ribbons, which is related to the difference of electronic conductivity in parallel and anti parallel alignment of magnetic moments, is up to 20\% for low doping level.  
\end{abstract}

\maketitle
\section{Introduction}
\begin{figure*}[!ht]
\includegraphics[width=1.0\linewidth]{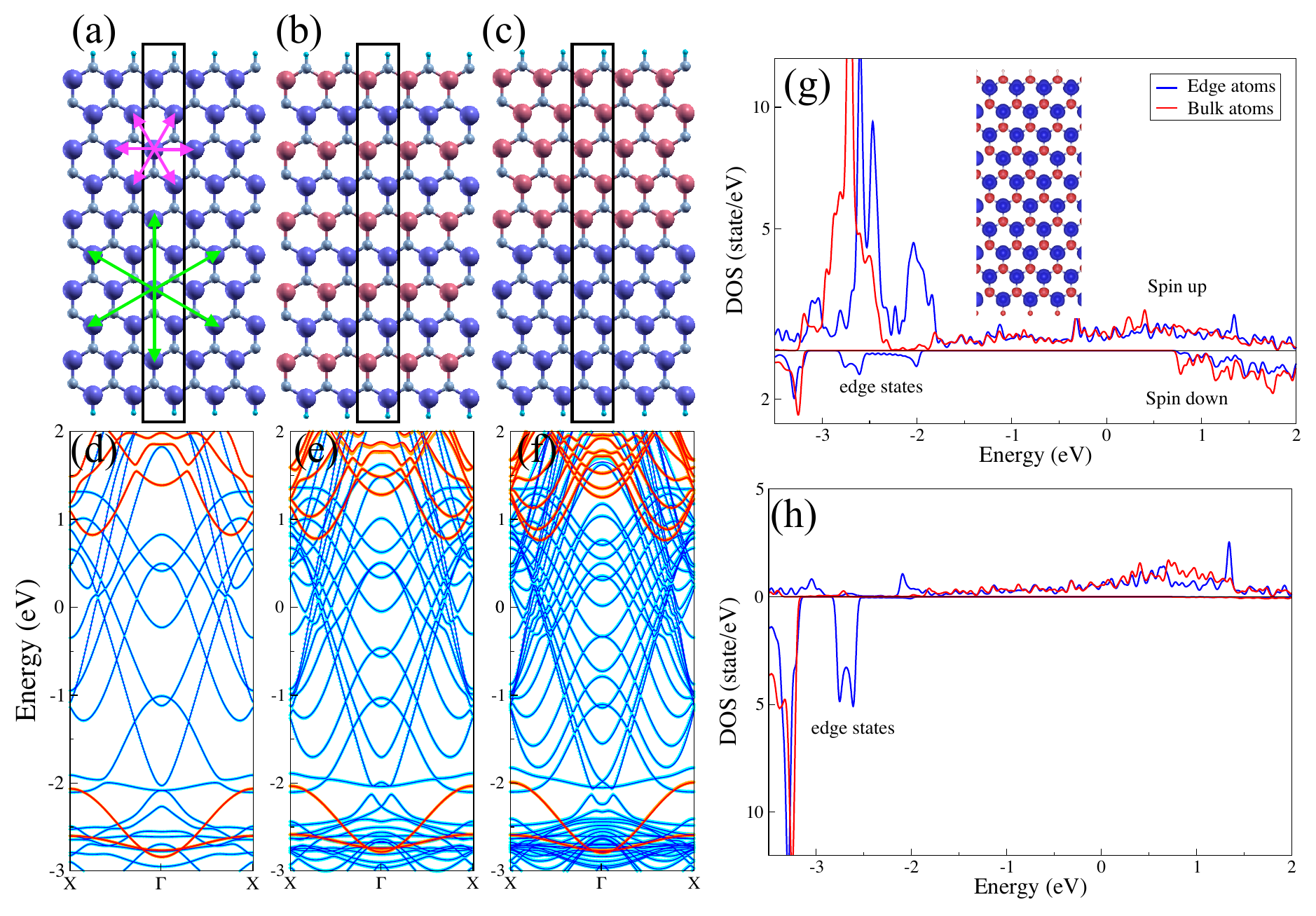}
\caption{The atomic and spin configuration for \ce{CrN} zigzag nanoribbons with $N=20$ and (a) FM, (b) AFM1, and (c) AFM2 spin configuration. Blue (red) Cr atoms have moments oriented in up (down) directions. The assumed ribbon supercell is marked by a black rectangle. Nearest and second-nearest metal neighbors are marked by purple and green arrows respectively. The electronic band structure for Z-\ce{CrN} nanoribbons of width (d) $N=8$, (e) $N=14$, and (f) $N=20$ atomic sites. The spin resolved DOS for bulk and edge states on (g) Cr, and (h) N atoms.}
\label{fig1}
\end{figure*}

In the past few years various two-dimensional monolayer materials, with different physical properties, such as metals, semiconductors, insulators, topological insulators, and ferroelectrics have been synthesized in laboratories and investigated theoretically. The low-dimensional magnetic order was not reported among physical properties until very recent discoveries. Besides the experimental observation of the ferromagnetic phase in \ce{CrI3} \cite{Huang_2018,Jiang_201, Zhong_20177, Liu_2019, Liu_2018, Jiang_20188}, \ce{VSe2} \cite{Bonilla_2018, Yu_2019}, few layer \ce{CrTe2} \cite{sun2019room}, \ce{MnSe2} \cite{O_Hara_2018}, and \ce{GdAg2} \cite{Ormaza_2016} at low temperature, the ferromagnetism (FM) has been proposed in many other 2D monolayers, including transition metal dichalcogenides, trihalides, and dihalides. The theoretically suggested FM monolayers include \ce{VS2} \cite{Luo_2017, Sun_2019}, \ce{VSSe} \cite{Zhang_2019}, \ce{CoBr2} \cite{Lv_2019, Chen_2017}, \ce{CrN} \cite{Modarresi_2019, Kuklin_2017}, \ce{CrGeS3} \cite{Ren_2019}, and MnX (where X=P, As, Sb)~\cite{Wang_2019}. For example, it was shown that \ce{CoBr2} is an intrinsic ferromagnetic semiconducting material with a Curie temperature of around 30~K \cite{Lv_2019, Chen_2017}. Also, hexagonal \ce{CrN} is a half metallic ferromagnetic 2D monolayer, with the Curie temperature around 200~K, which can be used in nanodevices with 100\% spin polarization abilities \cite{Modarresi_2019}. The magnetic orderings below the critical temperature in 2D monolayers are essential parts of many possible future technologies such as spin valves or magnetic data storage devices. 

However, in practice, all experimental samples have a finite size and are limited by edges. The edge effect is an interesting question toward the theoretical and experimental understanding of 2D ferromagnetic monolayers. The magnetic phase is a very well studied phenomenon for zigzag edges of graphene nanoribbon \cite{Modarresi_2017, G_l__2014}. The ferromagnetic and half metallicity are reported for hydrogen saturated \ce{InSe} nanoribbons \cite{Zhou_2018}. However, up to our best knowledge, the effect of edges has not yet been addressed for intrinsic FM monolayers. An important question is the effect of zigzag edges on the electronic and magnetic properties of nanoribbons. Nowadays, high performance computers with advanced computational packages are a standard tool for material investigation. In the present study, we address the electronic and finite-temperature magnetic properties of 1D zigzag edged \ce{CrN} (Z-\ce{CrN}) nanoribbons by using the Density-Functional-Theory (DFT) and Density Matrix Renormalization Group (DMRG) methods. We report the variation of the magnetic exchange interaction between the nearest and next nearest Cr atoms as a function of ribbon width. By using the calculated effective magnetic exchange between Cr atoms, the magnetic moment correlation at a finite temperature is estimated, proving the usefulness of this material for future nanodevices at high temperatures. Moreover, we investigate the spin-dependent transport through zigzag edged nanoribbons by using the Boltzmann transport theory and show possible perfect spin polarization for pure and doped \ce{CrN} zigzag nanoribbons. CrN ribbons have not yet been experimentally realized, however, there are experiments in which narrow ribbons have been synthesized from a variety of monolayer materials with a width of several dozen atomic layers~\cite{Lim2022, Aljarb2022}, even with a precisely controlled zigzag edge~\cite{Blackwell2021}.  

\section{Model and Methods}
\subsection{Electronic and magnetic structure}
First principle electronic calculations were performed based on the DFT code implemented in the Quantum ESPRESSO package \cite{QE-2017, QE-2009}. The ribbon's periodic direction is considered along the $x$ axis. The DFT supercell is marked by a black rectangle in Figs.~\ref{fig1}(a-c). Moreover, we use a 20~\si{\angstrom} vacuum spacing surrounding the supercell along the $y$ and $z$ directions to avoid interaction with the neighbor cells. A plane wave basis set is used with cut-off energy of nearly 1100 eV. The core electrons are modeled within the ultrasoft pseudopotential model. The exchange correlation between electrons is modeled by using generalized-gradient approximation with the Perdew-Burke-Ernzerhof functional \cite{PBE}. For the highly localized Cr-$d$ orbitals, we apply the Hubbard model with an effective on-site Coulomb repulsion for $d$ orbitals of Cr atoms. The Hubbard $U$ parameter is set to 3~eV as for the 2D \ce{CrN} monolayer \cite{Modarresi_2019,modarresi2020}. The integration over the first Brillouin zone (BZ) is performed by using the Monkhorst-Pack algorithm~\cite{Monkhorst_1976} with a $40 \times 1 \times 1$ $k$-point mesh. We define ribbon width $N$ (varying from 6 to 22) as the number of Cr and N atoms along the nonperiodic direction ($y$ axis). The zigzag edges of \ce{CrN} ribbon are passivated with H atoms to close the dangling bonds. All the analyzed \ce{CrN} nanoribbons are fully relaxed to find the minimum-energy atomic configuration. In the relaxed atomic configuration (see Fig.~\ref{fig1}(a-c)) the Hellmann-Feynman force acting on each atom is smaller than 0.002 eV/\si{\angstrom}. 

We performed the DFT+U calculations for three different magnetic configurations as presented in Fig.~\ref{fig1}(a-c), for a ribbon with $N=20$, by blue (red) colors denoting Cr atoms with spin aligned in the up (down) direction.
We select the spin configurations of Cr atoms that preserves translational symmetry along the ribbon: strictly polarized ferromagnetic (FM) state -- see Fig.~\ref{fig1}(a), and the two antiferromagnetic (AFM) states: one with collinear antiferromagnetic ordering between the nearest Cr neighbor atoms (AFM1) -- Fig.~\ref{fig1}(b), and the second with two antiparallelly oriented magnetic domains (AFM2) giving configuration where the moments at two zigzag edges are aligned oppositely -- Figs.~\ref{fig1}(c).

%The DFT+U calculations are performed for three different magnetic moment configurations. These three spin configurations on Cr atoms are presented in Fig.~\ref{fig1}(a-c), for a ribbon width of $N=20$, by blue (red) colors denoting Cr atoms with spin aligned in the up (down) direction. In the ferromagnetic (FM) state all magnetic moments on Cr atoms are oriented up, along the positive $z$ axis -- see Fig.~\ref{fig1}(a). We also consider two anti-ferromagnetic (AFM) magnetic moment configurations. In the AFM1 the nearest Cr atoms have opposite moment directions (Fig.~\ref{fig1}(b)), while for the AFM2 configuration the moments are aligned oppositely at two zigzag edges (Fig.~\ref{fig1}(c)) . 

The energy difference between the FM and the two AFM states is mapped into the Heisenberg spin model up to the second-nearest neighbor interaction (so-called $J_1$-$J_2$ model~\cite{Dagotto1989, Sirker2006}) as,
\begin{equation}
H=-\sum_{<i,j>}J_{1,ij} {\bf S}_{i}\cdot {\bf S}_{j}-\sum_{<<i,j>>}J_{2,ij} {\bf S}_{i}\cdot {\bf S}_{j} ,
\label{heis}
\end{equation}
where $J_{1(2)}$ is the exchange interaction strength between the nearest (next nearest) neighbor Cr atoms. 
The AFM1 state minimizes antiferromagnetic ordering for negative $J_2<0$, while the AFM2 is chosen just to have a second independent equation to determine the $J_1$ and $J_2$ integral values. The values of magnetic exchange interactions are obtained by comparing the energy difference of FM and the two AFM states in the DFT+U and the Heisenberg spin model giving
%energy in the Heisenberg model comprised with the DFT+U results \orange{[it is hard to understand this sentence, also, I rewrite equations below??]}:
%\begin{equation*}
%\begin{array}{l}
%J_1&=&(\Delta E^{AFM1-FM}-((N-2)/4)\Delta E^{AFM2-FM})/(2(N-2)),
%\\
%J_2&=&-(\Delta E^{AFM1-FM}-((N-2)/2)\Delta E^{AFM2-FM})/(4(N-2)),
%\label{J}
%\end{array}
%\end{equation*}
\begin{equation}
  \begin{aligned}
J_1=\frac{1}{2(N-2)}\left(\Delta E_1-\frac{N-2}{4}\Delta E_2\right),
\\
J_2=-\frac{1}{4(N-2)}\left(\Delta E_1-\frac{N-2}{2}\Delta E_2\right),
\label{J}
\end{aligned}
\end{equation}
where we define the energy differences as: $\Delta E_1= E^\mathrm{AFM1}-E^\mathrm{FM}$, and $\Delta E_2= E^\mathrm{AFM2}-E^\mathrm{FM}$.

The spin transport through Z-\ce{CrN} nanoribbon is calculated within the semiclassical Boltzmann transport theory by using the {\sc boltzwann} code \cite{Pizzi:196094} in connection with the {\sc wannier90} package \cite{MOSTOFI2008685}. The latter is used to interpolate the band structure by means of the maximally localized Wannier functions \cite{PhysRevB.56.12847, RevModPhys.84.1419}. The electrical conductivity along the $i$-direction is calculated as follows,
\begin{eqnarray}
	\sigma_{ii}\left (\mu,T \right ) = e^{2} \int_{-\infty}^{+\infty}  d\varepsilon \left (- \frac{\partial f \left ( \varepsilon, \mu, T\right)}{\partial \varepsilon}\right ) \Sigma_{ii} \left (\varepsilon\right), 
\end{eqnarray}
where $f \left ( \varepsilon, \mu, T\right)$ is the Fermi-Dirac distribution function, and $\Sigma_{ij}(\varepsilon)$ is the transport distribution function:
%\begin{eqnarray}
	%f \left ( \varepsilon, \mu, T\right) = \frac{1}{e^{\left (\varepsilon-\mu \right)/k_{B}T}+1 },
%\end{eqnarray}
\begin{eqnarray}
	\Sigma_{ij} \left (\varepsilon\right) =\frac{1}{V} \sum_{n,\mathbf{k}} v_{i}v_{j}\left(n,\mathbf{k}\right)  \tau\left(n,\mathbf{k}\right) \delta \left(\varepsilon-\varepsilon_{n,\mathbf{k}}\right).
\end{eqnarray}
Here, $\varepsilon_{n,\mathbf{k}}$ is the energy of the $n^{th}$ band at wave vector $\mathbf{k}$, $v_{i(j)}\left(n,\mathbf{k}\right)$ is the $i^{th}(j^{th})$ component of the group velocity, $V$ is the unit cell volume, and $\tau$ is the relaxation time. In all transport calculations, the relaxation time approximation is employed with the constant relaxation time set to $\tau=10$~fs, and the room temperature $T=300$~K.

\subsection{DMRG solution of 1D effective Heisenberg Hamiltonian}
\label{sec:dmrg}
The finite temperature correlation effects has been examined in the proposed effective Heisenberg spin model using the DMRG-like approach.
The DMRG method was originally developed for numerically finding the ground state of a one-dimensional gaped spin system with only nearest neighbor interactions \cite{White_1992, White_1993}.
The main idea of the DMRG-like approach is to restrict the Hilbert space to the subspace containing only a low-energy spectrum. It has been shown for 1D systems with local interactions that the wanted subspace is spanned by the eigenvectors corresponding to the most significant eigenvalues of the reduced density matrix \cite{Schollwock.2011, Eisert.Cramer.2010}.
Let us assume that the system is spatially divided into two subsystems. The reduced density matrix is obtained after a partial trace over one of the subsystems from the full density matrix. The eigenvalues of such a system are proportional to the entanglement entropy between two parts of that system. DMRG is such a powerful method for these kinds of systems because the ground-state entanglement entropy, associated with the eigenvalues of the reduced density matrix, increases proportionally to the edge of the system, what is called area law \cite{Eisert.Cramer.2010}. 
For a 1D system, the edges of the system consist of just two ends of the chain, so its area is constant, which implies that the dimension of the low-energy subspace is constant.

The modern approach to the DMRG method, based on the matrix product states (MPSs) and matrix product operators (MPOs), allows simulating two dimensional structures like CrI$_3$ \cite{Phys.E.115520} and real time and imaginary time evolution of quantum states \cite{White_Feiguin_2004, Feiguin_White_2005, Schollwock.2011}. The dimension of MPSs and MPOs is reduced in the same way as in the original DMRG method. One of the methods to examine the finite-temperature properties of the quantum system is a thermalization of a quantum state from the infinite temperature by the application of the operator in the form $e^{\frac{-H}{k_B T}}$, where $k_B$ is the Boltzmann constant. To use this method, the original studied system has to be in contact with some thermal bath. It can be effectively done by the DMRG method in MPSs and MPOs formulation. The original system is enlarged by its copy, which acts as a thermal bath \cite{Feiguin_White_2005, Schollwock.2011}.
Then, the state vector $\left| \psi \right>$ has the form of the tensor product of the real and auxiliary sites (called ancillas). Let us denote the energy eigenstates of the real system by the ${\left| n \right>}$, and correspond to them eigenstates of the auxiliary systems by ${\left| \Tilde{n} \right>}$. Then, the unnormalized vector state in a non-zero temperature $T$ can be defined in the following way,
\begin{equation}
    \left| \psi(T) \right> = e^{\frac{-H}{2k_B T}} \left| \psi(\infty) \right> = \sum_{n} e^\frac{-E_n}{2 k_B T} \left| n \Tilde{n} \right>,
\end{equation}
and $\left| \psi(\infty) \right> = \sum_n \left| n \Tilde{n} \right> $ corresponds to the state in the infinity temperature. It is important to note that the Hamiltonian $H$ acts only on the physical part of the vector states and ancillas evolve only by the entanglement with the real sites. Therefore, entanglement between real and auxiliary sites is crucial and instead of choosing the vector state for the infinity temperate in the form $\left| \psi(\infty) \right> = \sum_n \left| n \Tilde{n} \right> $ it is more accurate to choose it as $\left| \psi(\infty) \right> = \sum_s \left| s \Tilde{s} \right>$, where $s$ represents the real spin site, and  $\Tilde{s}$ corresponds to auxiliary ones. Now, the average value of any operator $O$ in temperature $T$ could be calculated in the following simple way,
\begin{equation}
    \left<O\right> = \frac{\left< \psi(T) \right| O \left| \psi(T) \right>}{\left< \psi(T)|\psi(T) \right>}.
\end{equation}
We have used this approach to examine the thermodynamic properties of the proposed spin chain model (\ref{heis}). The evolution in the imaginary time has been done iteratively by applying the evolution operator $U(-\Delta \beta)$ acting on the state vector,
\begin{equation}
    \left| \psi\left(T-\frac{1}{k_B \Delta \beta}\right) \right> = U(-\Delta \beta) \left| \psi(T) \right> = e^{\frac{-H \Delta \beta}{2} \left|\psi(T) \right>}.
\end{equation}
The effective approach for the system with not only nearest neighbor interactions has also been used for the operator $U$~\cite{Zaletel_Mong_2015}. Then, for the state in temperature $T$, the correlation function $C(r_{ij}) = \mathbf{S_i} \cdot \mathbf{S_j}$ has to be calculated. The correlation length $\epsilon$, which is crucial for our consideration, has been calculated by simply fitting an exponential function to the correlation function $C(r_{ij}) = A e^\frac{-|r_i-r_j|}{\epsilon}$, where $r_i$ and $r_j$ are the positions of $i$ and $j$ sites. 

In the Supplementary Information, we analyze the scaling of the correlation length $\epsilon$ with the spin chain length and maximum MPS dimension. After this analysis, the latter calculations have been performed for the chain of $L=120$ spins and the maximum MPS dimension $\chi=600$. The length of such a chain is equal to $207.6$ in the units of the Cr-N bond. In the DMRG simulations, we assumed quite long chains ($L=120$) to correctly capture the spin correlation length effect, which might be much longer than the ribbon width $N$. Our calculations have been performed by using the open-source library for DMRG simulations, TeNPy \cite{tenpy}.

\section{Results and Discussion}
\subsection{Electronic structure and finite temperature magnetism}
\begin{figure*}[!ht]
\includegraphics[width=1.0\linewidth]{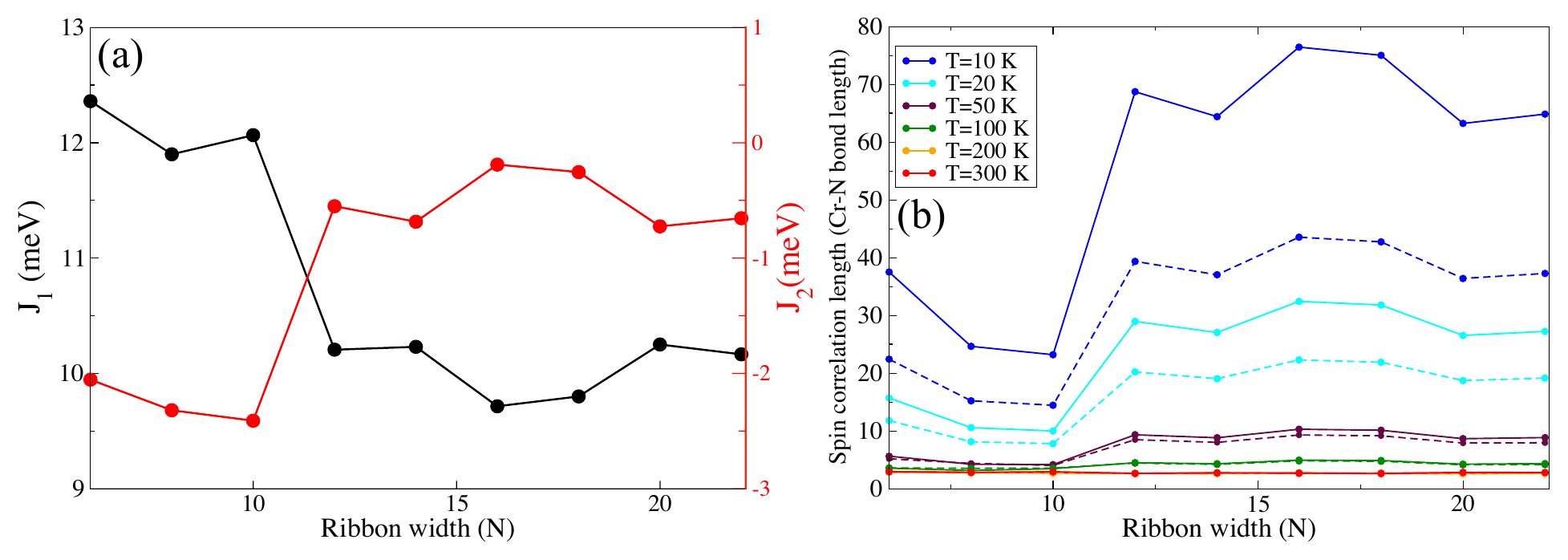}
\caption{(a) Width-dependent exchange interaction between nearest and next nearest neighbor Cr atoms. (b) The spin correlation length as a function of ribbon width at different temperatures. The solid and dash lines represent the inclusion and exclusion of magnetic anisotropy in the DMRG analysis, respectively.}
\label{fig2}
\end{figure*}

\begin{figure*}[t]
\includegraphics[width=1.0\linewidth]{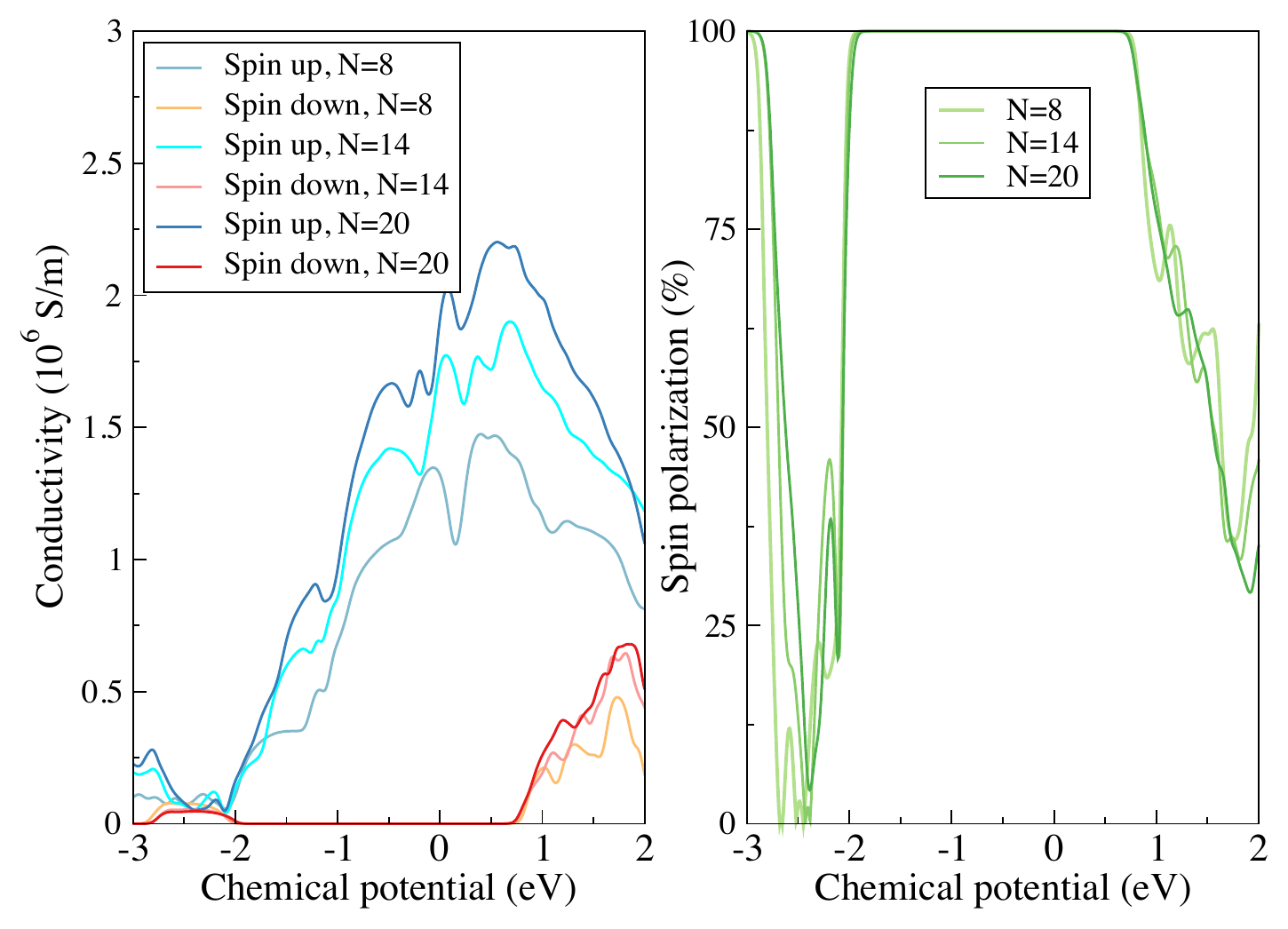}
\caption{(a) Spin-dependent electronic conductivity, and (b) spin polarization ratio as a function of chemical potential for Z-\ce{CrN} nanoribbons with $N=8$, 14 and 20.}
\label{fig3}
\end{figure*}
%The armchair nano ribbons there are odd number of Cr atoms per unit cell. We construct the non collinear FM and AFM spin states as shwn 

%The formation energy of zigzag and armchair edges are calculated for narrow ribbons by considering total energy for ribbons and bulk 2D monolayer. The formation energy is defined as the energy difference between ribbon and 2D-monolayer multiplied by number of Cr-N pairs and H-atoms energy. For example for the zigzag edge the formation energy for each edge atom will be $E^{formation}=\frac{1}{2}\left [ E^{ribbon}-(N/2)E^{2D-\ce{CrN}}-2E^H \right ]$. According to our calculations, zigzag \ce{CrN} nanoribbons has lower edge formation energy and are more stable which is consistent with previous reports for CrI$_3$ ribbons \cite{Wang_2019}. The energy difference with armchair ribbons is around 100 meV for narrow ribbons. We will focus on the nanoribbons with zigzag edges of hexagonal \ce{CrN}.  
Two types of nanoribbons can be constructed by cutting the hexagonal \ce{CrN} monolayer, namely with the zigzag and armchair edges. Typically, the highly reactive dangling edge bonds are saturated by hydrogen atoms (see Fig.~\ref{fig1}). In the case of graphene, there is an antiferromagnetic spin order on the two opposite zigzag edges of a ribbon. Here, for the sake of comparison of magnetic order, we chose to focus only on zigzag (Z-) edges in hexagonal \ce{CrN}. We compare the electronic energy of FM, AFM1, and AFM2 configurations for Z-\ce{CrN} ribbons, and find out that the FM magnetic phase is the ground state for all the zigzag ribbon widths. Among anti-ferromagnetic phases, the AFM2 has lower energy, which is a similar result to the ground state of zigzag graphene nanoribbons \cite{grapheheribbons}. The average Cr-N bond length is 1.87~\si{\angstrom}, which is close to the bond length in a 2D hexagonal \ce{CrN} monolayer \cite{Kuklin_2017}. The electronic band structure of Z-\ce{CrN} nanoribbons for $N=8$, 14, and 20 are presented in Fig.~\ref{fig1}(d-f). The Fermi level crosses spin up electronic states, while there is an energy gap for spin down states. The valence band maxima for spin down are at the edge of the first BZ. The DFT electronic bands are successfully interpolated using the maximally localized Wannier functions. The Wannier model is also used to calculate the spin dependent electronic conductivity. The net magnetic moment in the FM phase Cr atom is 3~$\mu_B$ per Cr atom which is equal to the 2D monolayer \ce{CrN} \cite{Modarresi_2019, Kuklin_2017}. 
The electronic bands of FM, AFM1, and AFM2 for N=20 are compared in the Supplementary Information (See Fig.~S2). To observe the effect of localized edge states, the projected density of states, on the edges, and for bulk Cr and N atoms, are plotted in Fig.~\ref{fig1}(g-h). For both Cr and N, the edge and bulk atoms contribute almost equally to spin up states around the Fermi level aligned to 0 energy. But for spin down states, only edge atoms contribute to localized bands between $-3$ to $-2$~eV. These localized bands arise from both Cr and N atoms. The energy band gap between spin down states is 2.9, 2.8, and 2.76~eV for $N=8$, 14 and 20, respectively (see Fig.~\ref{fig1}(d-f)). The edge states on Cr and N atoms reduce the energy band gap for spin down states with respect to the 4~eV for 2D monolayer \ce{CrN} \cite{Modarresi_2019, Kuklin_2017}. The inset in Fig.~\ref{fig1}(g) presents calculated spin polarization for the FM ground state in Z-\ce{CrN} nanoribbon with $N=20$. The N atoms gain a net magnetic moment antiparallel to the Cr moments. The half metallicity feature of monolayer \ce{CrN} is conserved in Z-\ce{CrN} nanoribbons which motivates its future spintronic applications. 

In the following sections, we will study the spin correlation length at finite temperature and spin dependent transport properties of Z-\ce{CrN} nanoribbons. The energy difference between FM and AFM phases is mapped into the Heisenberg spin model with the nearest and next nearest neighbor exchange interactions between Cr atoms. 
The Cr magnetic moments induce a net magnetic moment in the N atoms (approximately $-0.2$ Bohr magneton/atom) which is neglected in the DMRG analysis. 
The exchange interaction $J_{1(2)}$ is plotted as a function of ribbon width in Fig.~\ref{fig2}(a). The positive value of $J_1$ and negative for $J_2$ confirm the parallel and anti-parallel coupling between the nearest and next nearest Cr atoms. The value of $J_1$ parameter is nearly independent of the ribbon width and varies between 10 and 12~meV for ribbons with $N=6$ and $N=20$. The value of $J_2$ parameter is around $-2.5$~meV for narrow ribbons and goes to zero for wider Z-\ce{CrN} nanoribbons. The $J_1$ value is larger than the $J_2$ absolute value by almost one order of magnitude for wider nanoribbons.

The Mermin-Wagner-Hohenberg theorem prohibits the long-range magnetic order at finite temperature for the isotropic Heisenberg Hamiltonian in low dimensional systems ($D\leq2$) \cite{Mermin_1966, Hohenberg_1967}. It was shown that magnetic anisotropy energy stabilizes finite-temperature long-range magnetic order in 2D monolayers~\cite{Modarresi_2019, memarzadeh, modarresi2020}. However, in 1D, the spontaneous formation of FM domains, separated by kinks, avoids the magnetic long-range order in both Ising and Heisenberg spin models~\cite{majlis2007}. Consequently, there is a finite-range magnetic correlation between localized magnetic moments in the \ce{CrN} nanoribbons. In Fig.~\ref{fig2}(b), the spin correlation length is plotted as a function of ribbon width in the unit of Cr-N atomic bond length with the inclusion (solid line) and exclusion (dash line) of magnetic anisotropy. Here the value of single ion magnetic anisotropy is set to 0.73 meV equal to the monolayer value \cite{Modarresi_2019}. The width dependency arises from different values of $J_1$ and $J_2$ for narrow and wide nanoribbons. The spin correlation is mainly determined by the value of the next nearest neighbor exchange. At sufficiently low temperatures, for wide ribbons, the spin correlation length is much longer than the Cr-N bond length which is desired for realistic 1D nanodevices. By increasing temperature, the spin correlation reduces to only one Cr-N bond length which shows the absence of local 1D magnetic order at high temperatures. 
The effect of single ion magnetic anisotropy energy depends on the temperature. At a very low temperature ($T=10$~K), the correlation length is doubled by the inclusion of the small magnetic anisotropy energy. By increasing temperature the effect of magnetic anisotropy energy on the correlation length disappeared which is related to the thermal fluctuations. 
Although the inclusion of magnetic anisotropy energy increases the spin correlation length at low temperatures, it does not produce 1D long range magnetic order. 
The Curie temperature for 2D \ce{CrN} monolayer is estimated 1084~K~\cite{Modarresi_2019} but the magnetic order at zigzag edges completely disappeared above 100 K. The effective Heisenberg Hamiltonian and the absence of long range magnetic order at the edges can be valid for all 2D magnetic monolayers. According to our result, the experimental observation of macroscopic magnetization at 2D requires high quality surfaces with large grain boundaries.

%Temt nge magnetic long 2D temperature is an important factor for low dimensional magnetic materials since the prediction of absence of long-range magnetic order at any finite temperature \cite{Mermin_1966,Hohenberg_1967}.
%The range of coupling between spins is characterized by spin correlation length $\zeta$ which determined the exponential decay of correlation between magnetic moments. The spin correlation length is estimated from the $J_1$ value in the absence of external magnetic field for the 1D Ising model \cite{Zhou_2018},
%\begin{equation}
%\zeta =\left [\mathrm{ln}(\mathrm{tanh}(J_1/K_BT)) \right ]^{-1}.
%\end{equation}

%This is still possible operating nano-devices based on narrow ribbons at temperature of liquid nitrogen. According to our results at T=77K for ribbon with width lower than 2 nm the spin correlation length is larger than ribbon width.

%To understand the role of different edge passivation, we also investigate the effect of halides edge passivation on the value of $J_1(2)$ in Z-\ce{CrN} nanoribbons. According to results, the value of $J_1$ can change between 9.8 to 12.7 eV for passivation with F and Br atoms. Also the absolute value of $J_2$ increases from 0.5 to 2.8 eV. The engineering of exchange coupling can increase the spin-correlation at higher temperatures. In the following part we will study transport characteristic of Z-\ce{CrN} nanoribbons as a possible nano device. 

\subsection{Spin-dependent transport}
To observe the effect of magnetic moments on the transport properties of Z-\ce{CrN} nanoribbons, we also calculate the spin-resolved conductivity as a function of chemical potential. Here the change in chemical potential can be interpreted as $n$ and $p$ type doping. The spin-dependent conductivity for different ribbon widths is plotted in Fig.~\ref{fig3}(a). The half metallicity of Z-\ce{CrN} nanoribbons is reflected as pure spin up polarization at the small value of chemical potential. 
The value of relaxation time $\tau$ can change spin dependent conductivity in Fig.~\ref{fig3}(a). But it does not change our main conclusion for the pure spin current at the low value of chemical potential. Also one can show that in the absence of heavy atom impurities, the spin-flip scattering rate is several orders of magnitude smaller than the spin transport scattering rate \cite{Modarresi_2019}. The conductivity weakly depends on the ribbon width. For example, the ribbon with $N=14$ is nearly twice as wide as the one with $N=8$, while the conductivity at $\mu=0$ is almost equal for these two nanoribbons. The spin polarization is defined as the ratio of spin up and spin down differences to the total conductivity, $\mathrm{SP}=\left [ \frac{{\sigma^\uparrow-\sigma^\downarrow}}{\sigma^\uparrow+\sigma^\downarrow} \right ]\times100$. The spin polarization for $N=$ 8, 14, and 20 as a function of chemical potential is plotted in Fig.~\ref{fig3}(b). For pure ($\mu=0$) and $p$-doped ($-2<\mu<0$) Z-\ce{CrN} nanoribbons, the only contribution to spin conductivity arises from spin up states and spin polarization reaches 100\%, which is motivating for more theoretical and experimental investigations. Also on the other hand, for $n$-doped nanoribbons, the spin polarization is reduced sharply for $\mu>0.5$. When increasing the $n$-doped level and further increasing chemical potential, the conductivity is reduced which is not desired for spintronic devices. In General, the Z-\ce{CrN} nanoribbons are functioning as spin polarizers sensitive to the doping level.

\section{Conclusion}
In summary, we study the electronic, magnetic, and electronic transport properties of zigzag \ce{CrN} nanoribbons within the combination of DFT and DMRG frameworks. The zigzag \ce{CrN} nanoribbons are in the FM ground state with a net magnetic moment of 3 $\mu_B$ per Cr atom. The nearest Cr atoms are coupled in parallel with $J_1$ equals 10 to 12~meV, while the exchange with the next nearest neighbor is anti-parallel coupling with $J_2$ equals $-2$ to $-0.5$~meV. According to the spin dependent transport calculations, the pure and $p$-doped Z-\ce{CrN} nanoribbons are potentially perfect spin polarizers, suitable for future spintronic applications. However, the DMRG analysis of spin-spin correlation length shows that the possible technological application of Z-\ce{CrN} nanoribbons is limited to narrow ribbons and at relatively low-temperature regimes. 
%Finally, we have propose a nano spin valve device based on two Z-\ce{CrN} electrodes separated by a boron phosphide ribbon layer. The electronic conductivity depends on the relative direction of magnetic moments and we report considerable magnetoresistance up to 20\%.

\section*{Acknowledgments}
Our calculations were performed at the Wroc\l{}aw Center for Networking and Supercomputing (WCSS). The DMRG calculations were performed using the TeNPy library (version 0.9.0)~\cite{tenpy}. M.~K. was supported by the National Science Centre (NCN, Poland) under the grant: 2019/33/N/ST3/03137. 
J.~P. acknowledges support from National Science Centre, Poland, under grant 2021/43/D/ST3/01989.
This work was supported by the Scientific and Technological Research Council of Turkey (TUBITAK) under Project No. 119F361. The numerical calculations reported in this paper were partially performed at TUBITAK ULAKBIM, High Performance and Grid Computing Center (TRUBA resources).

\section*{References}
\bibliography{main}

%\appendix
%\twocolumngrid
%\section*{Appendix}
%\subsection*{DMRG parameters optimization}
%\label{app:dmrg}
%\textbf{The DMRG method described in section \ref{sec:dmrg} requires selecting the optimal set of parameters, like maximal MPS dimension $\chi$ and the chain length $L$. The optimal parameters have been found by solving the Hamiltonian (\ref{heis}) for the ribbon of width $N=16$ with magnetic anisotropy in the temperature $T=10$~K, because, as shown in Fig.~\ref{fig2} (b), the most correlated state is expected in that case. The correlation length $\epsilon$ in the function of chain length $L$ is shown in Fig.~\ref{fig:app_dmrg} (a) for fixed $\chi = 600$. Similarly, in Fig.~\ref{fig:app_dmrg} (b) are shown results for fixed $L=80$ depending on MPS dimension $\chi$. Finally, by analyzing both plots, the $L=120$ and $\chi=600$ have been chosen.}

%\begin{figure}[h]
%\includegraphics[width=1.0\linewidth]{stability.pdf}
%\caption{Correlation length in the ribbon of width $N=16$ for $T=10$~K, (a) as a function of length $L$ for $\chi=600$, and (b) as a function of MPS bond dimension for $L=80$. }
%\label{fig:app_dmrg}
%\end{figure}

\end{document}